# Programmable Broadband Polarization-Preserving Optical Delay Line with Faraday Mirrors

Salem F. Hegazy, *Member, IEEE,* Salah S. A. Obayya, *Fellow, IEEE, Fellow, OSA,* and Bahaa E. A. Saleh, *Life Fellow, IEEE, Fellow, OSA*

***Abstract*—We report a broadband optical delay line programmable for delays up to 1 μs with nanosecond resolution, while preserving an *arbitrary* and *unknown* polarization state of the original light. The delay line comprises bidirectional pass in cascaded segments of standard single-mode fibers, along with terminal reflection from Faraday mirrors (FMs) at both ends. Each segment contains either a short path or a long path, selected by an optical switch. The FMs auto-compensate for random polarization fluctuations introduced by the optical fibers, thereby restoring the initial polarization state. Broadband operation of the delay line was assessed by investigating dispersion of both the fiber and the FMs. Surprisingly, we found that the two-pass propagation in the fiber can mitigate dispersive distortion introduced by the FM. Numerical experiments demonstrated that the output polarization state of the delay line has a mean fidelity better than 99% over a range of 50-nm deviation from the central wavelength. Remarkably, the mean fidelity often surpasses that in the absence of the fiber itself.**

## I. Introduction

OPTICAL delay lines are designed to temporarily store optical signals with minimal loss and distortion. They play a crucial role in various applications of all-optical switching and signal processing [1-3]. They are essential in optical communications for optical packet synchronization and buffering, and also for simplifying several communication schemes, including rate-adjustable phase-shift keying (PSK) [4] and hybrid differential PSK-multipulse position modulation (MPPM) [5-8]. Programmable optical delay lines can also stimulate new areas of research in delay-sensitive optical systems, such as optical-chaos systems based on delayed feedback [9,10].

The delay and storage of photons with arbitrary and unknown polarization states is important in several quantum information and quantum computation applications. It is useful in switching, buffering, and routing of polarization-coded quantum signals [11] and polarization-entangled photons [12,13]. Additionally, it can be integrated into quantum key distribution systems [14-17] to scramble the critical timing information.

Optical delays are generally realized using two approaches. The first offers continuous delay tuning by control of group velocity in optical structures such as Bragg gratings [18], photonic-crystal waveguides [19-21], and resonators [22,23]. Resonators are preferred due to their compact size and simple structure [24,25]; their group delay is varied by detuning the resonance wavelength. However, there is always a trade-off between optical bandwidth, group velocity dispersion, and maximum achievable delay [26]. As a result, this approach is suitable for achieving relatively short delays, limited by the delay-bandwidth product.

The second approach for implementing optical delay involves modifying routing paths through serial or parallel line topologies using $2 \times 2$ or $N \times N$ optical switches [27, 28]. These switchable delay lines can be realized using both fiber-optic components and integrated photonics [29].

Ideally, the output of a delay line is a lagging replica of its input. However, in practice, because of thermal and mechanical variations, optical fibers induce slow drifts and fluctuations in phase and polarization, altering the polarization state of the output signal. Therefore, while fiber-optic implementations can achieve much longer delays compared to integrated photonics, they often suffer from poor stability [30]. Also, delaying the optical signal while preserving its arbitrary polarization state cannot be accomplished by use of a polarization-maintaining fiber, which preserves only the polarization of its two principal

This work was supported by the Academy of Scientific Research and Technology (ASRT) through JESOR Program under Grant 5283. (Corresponding authors: Salem Hegazy and Bahaa Saleh.)

Salem F. Hegazy is with the National Institute of Laser Enhanced Sciences, Cairo University, Giza 12613, Egypt (e-mail: salem@niles.cu.edu.eg).

Salah S. A. Obayya is with the Center for Photonics and Smart Materials, Zewail City of Science and Technology, Giza 12578, Egypt (e-mail: sobayya@zewailcity.edu.eg).

Bahaa E. A. Saleh is with CREOL, The College of Optics and Photonics, University of Central Florida, Orlando, FL 32816 USA (e-mail: besaleh@creol.ucf.edu).

states.

Since the concept was introduced by Martinelli in 1989 [31,32], the Faraday mirror (FM) has been extensively used in compensating polarization fluctuations in retracing optical fibers. This approach has been utilized in multiple applications [33]. For instance, it has found application in stabilizing self-tuning lasers [34, 35], fiber-optic interferometers [36], and quantum key distribution [14-17]. Additionally, it has been employed to ensure highly accurate polarization-based switching in optical delay lines using non-polarization-maintaining optical fibers [37, 38]. The main feature of the FM is that its output polarization state is always orthogonal to the input state. Retracing an optical fiber after reflection from an FM in an orthogonal polarization state passively and automatically compensates all polarization distortions encountered. This occurs at the central wavelength of the FM operation. However, despite the wide operational bandwidth (~100 nm) of most commercially available FMs, the literature lacks adequate investigation into the impact of FM dispersion on the output polarization state.

This paper presents a broadband fiber-optic delay line with delays in the microsecond range that preserve the polarization state of the delayed signal. The delay line comprises a number of binary single-mode fiber-optic stages and two terminal Faraday mirrors. Thanks to the FMs, bidirectional light traveling through the fiber stages always yields a polarization state orthogonal to the input state for any programmed delay value. This cancels out all polarization fluctuations encountered by fiber propagation. The FM placed at the other end of the delay line effectively restores the polarization of the input signal regardless of the state of polarization of the input. This occurs precisely at the central wavelength of operation.

Performance under broadband operation at the edge of the operational range of the FM is assessed. This includes the impact of chromatic dispersion of the fiber and the dispersive polarization rotation introduced by the FMs. We carry out numerical experiments that show possible output polarization states for an off-band light pulses (50-nm from the FM central-wavelength). Experimental results demonstrated the interplay between the random polarization transformation of the optical fiber and the dispersion of the FM. We show that the randomness of the fiber transformation reduces the effect of FM dispersion on the fidelity of the output state of polarization.

## II. Programmable Delay System

A schematic of the multi-stage delay system is depicted in Fig. 1. Consider a light pulse fed into a series of $K$ delay stages separated by $2 \times 2$ optical switches (based, e.g., on a Mach-Zehnder interferometer configuration). In the first stage, the optical switch (OS) routes the light pulse either to a short fiber segment of length $L_o$ or to a long one of length $L_0 + L$. The short and long arms are then connected to another OS, leading to the next stage. The first delay stage thus switches between two delay times:

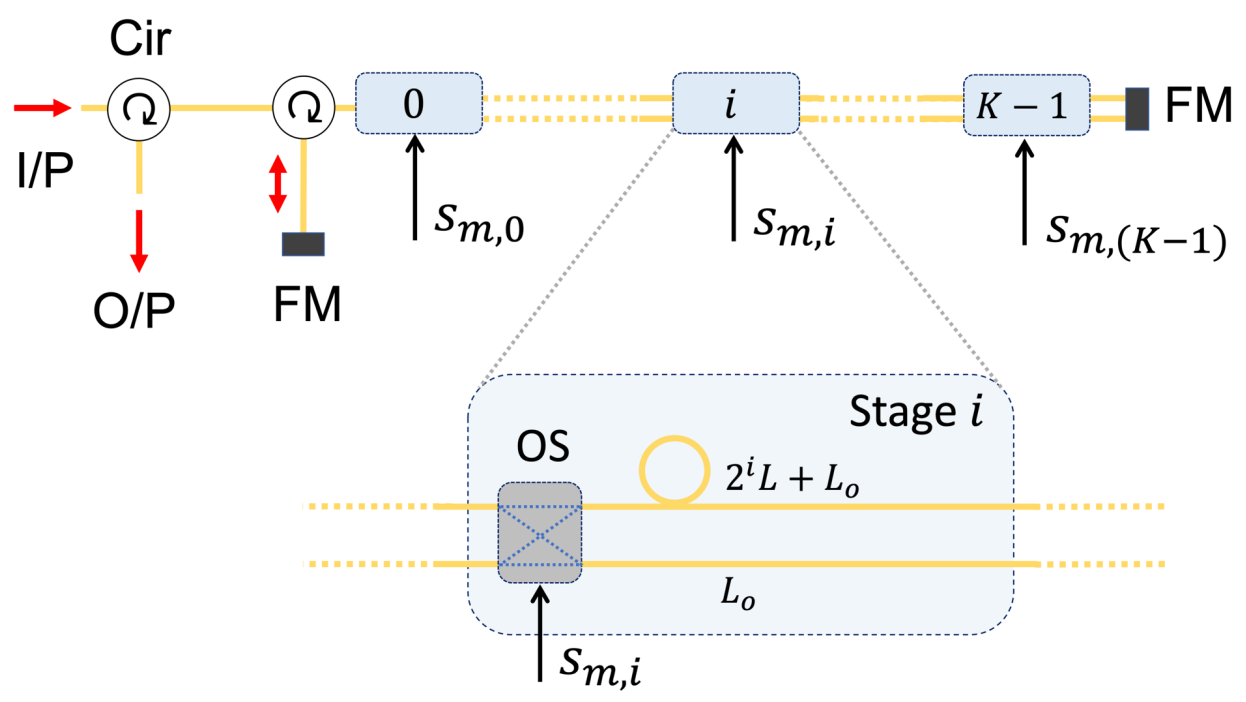

Fig. 1. Schematic of $K$–stage programmable polarization-preserving delay line using single-mode optical fibers. In every stage $i$, an optical switch (OS) routes the signal either to a long fiber of length $2^i L + L_o$ or to a short one of length $L_o$. After a forward pass through the sequence of fibers, a Faraday mirror (FM) reflects the light back, converting it into an orthogonal state of polarization (SOP) before its backward propagation through the same fiber. Before exiting the delay line, another FM restores the initial SOP of the optical signal.

$$\tau + \tau_o = \frac{1}{\mathrm{v}_g}(L + L_o)\,, \quad \tau_o = \frac{L_o}{\mathrm{v}_g}\,, \tag{1}$$

where $\mathrm{v}_g$ is the effective group velocity along the optical fiber.

Along the series of $K$ delay stages, each stage $i$ ($i \in \{0,1,\ldots,K-1\}$) has two arms: a short fiber segment of length $L_o$, and a long one of length $2^i L + L_o$. After the first pass through the $K$ stages, a Faraday mirror (FM) reflects the light pulse back through the system. The FM reflected the light in an orthogonal polarization. As shown in the next section, this passively and automatically compensates all polarization fluctuations experienced by the light pulse throughout the round trip [31,32] (this assumes that the time constant of the fluctuations is much less than the round trip time, which is typically the case). Before exiting the system, the optical signal is routed by means of an optical circulator (Cir) to a second FM in another path that restores the initial polarization.

The control signals of the optical switches along the $K$ stages can thus govern the round-trip time of the light pulse as

$$T_m = 2\sum_{i=0}^{K-1} S_{m,i}\left(2^i\tau + \tau_o\right) + \bar{S}_{m,i}\,\tau_o\,, \tag{2}$$

where the first and second terms are associated with propagation in the long and short arms, respectively, the subscript $m$ denotes the $m$th delay step ($m \in \{1,2,\ldots,2^K\}$), $S_{m,i}$ is the OS control signal in the $i$th stage $\left(S_{m,i} \in \{0,1\}\right)$ that realizes the delay $T_m$, $\bar{S}_{m,i}$ is NOT$\left(S_{m,i}\right)$, and the factor of 2 is attributed to the two-pass propagation. Equation (2) shows that, based on the values of the control signals $S_{m,i}$, the traverse time $T_m$ may take one of the values $\{0, 2\tau, 4\tau, 6\tau, \ldots, (2^K-1)2\tau\}$, up to a fixed delay offset of $2K\tau_o$. The maximum delay, for which, $S_{2^K,i} = 1\ \forall\ i$, can be expressed as

$$T_{2^K} = 2K\tau_o + \sum_{i=0}^{K-1} 2^{i+1}\tau = 2K\tau_o + (2^K - 1)2\tau. \quad (3)$$

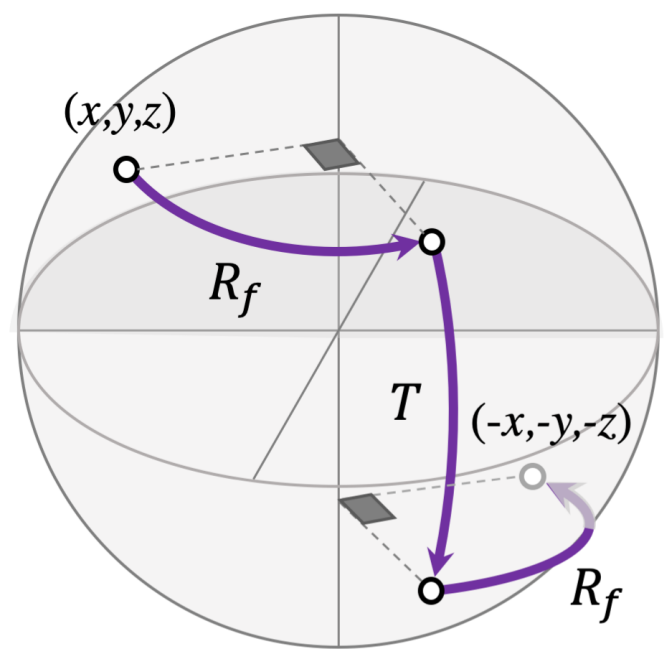

Fig. 2. Evolution of a polarization state on the Poincaré sphere due to reflection by a Faraday mirror (FM). This operation involves a forward pass through a $\lambda/4$ Faraday rotator (denoted by the operator $R_f$), reflection by a mirror (operation $T$), and backward pass through the Faraday rotator (operator $R_f$ again). The Faraday rotator applies a nonreciprocal 90-degrees rotation around the north-south axis (z axis) in both forward and backward propagation. The mirror reflection is described by a symmetry through the equatorial plane. The transformation $R_f T R_f$ converts a polarization state of coordinates $(x, y, z)$ to its orthogonal state $(-x, -y, -z)$.

## III. Passive Autocompensation of Polarization Fluctuations Using Faraday Mirror

The optical signal after the first pass through the fiber and just before the FM is assumed to be in the polarization state

$$\alpha|R\rangle + \beta \exp(i\phi)\,|L\rangle, \quad (4)$$

where $|R\rangle$ and $|L\rangle$ are basis states formed by right- and left-circular polarization, $\alpha, \beta$ are real numbers satisfying the normalization condition of the state vector, and $\phi$ is a relative phase. In (4), the polarization state vector is represented by a ket $|.\rangle$, following Dirac notations. The FM consists of a $\lambda/4$ Faraday rotator and a terminal mirror in front. As depicted in Fig. 2, action of the Faraday rotator can be shown on the Poincaré sphere by a 90° rotation about the $z$ axis. This operation can be expressed mathematically in the *R/L* basis by the operator

$$R_f = \exp(-i\pi\sigma_z/4) = \begin{bmatrix} \exp(-i\pi/4) & 0 \\ 0 & \exp(i\pi/4) \end{bmatrix}, \quad (5)$$

where $\sigma_z$ is the Pauli $Z$ matrix. This produces the state

$$\alpha|R\rangle + \beta \exp i(\phi + \pi/2)\,|L\rangle.$$

Action of the mirror on the polarization state is described by a symmetrical move through the equatorial plane. That is, the state $|R\rangle$ is exchanged with $|L\rangle$, and direction of rotation of any elliptical polarization state is reversed, while the ratio of the major and minor axes is kept unchanged, and linear polarization is also kept unchanged. This action can be expressed mathematically by

$$T\colon \begin{pmatrix} \alpha \\ \beta \exp i(\phi + \pi/2) \end{pmatrix} \rightarrow \begin{pmatrix} \beta \\ \alpha \exp i(\phi + \pi/2) \end{pmatrix}, \quad (6)$$

which involves interchanging and conjugating the two factors of the state. This is not a unitary operation (which is not a problem [14]), as it involves switching from the right-handed reference $x, y, z$ to the left-handed reference $x, y, -z$.

The Faraday rotator is a nonreciprocal component. Its action is described by the same operator $R_f$ in either directions of propagation. Therefore, upon passing through it in the reverse direction, the operator $R_f$ is applied twice. The overall action of the Faraday mirror on a state of Cartesian coordinates $(x, y, z)$ on the Poincaré sphere is thus described by

$$(x, y, z) \rightarrow (-x, -y, -z),$$

where the negative sign in the $x, y$ coordinates is due to the overall 180° rotation about the *z* axis by two passes through the Faraday rotator, and the negative sign in the *z* coordinate is due to reflection from the mirror. This converts any polarization state to its orthogonal state, which can be expressed by

$$R_f T R_f\colon \begin{pmatrix} \alpha \\ \beta \exp(i\phi) \end{pmatrix} \rightarrow \begin{pmatrix} \beta \\ -\alpha \exp(i\phi) \end{pmatrix}. \quad (7)$$

The fiber-optic line can be modeled as a number of birefringent elements. Each element represents a fiber segment applying a unitary operation on the polarization state. The total effect of a two-pass propagation through the delay line involving the Faraday-mirror reflection can be described collectively by the operator

$$U_{\text{fiber}}^{-1} R_f T R_f\, U_{\text{fiber}} = U_1^{-1} U_2^{-1} \ldots U_M^{-1}\, R_f T R_f\, U_M \ldots U_2 U_1, \quad (8)$$

which equals $R_f T R_f$ for any number of fiber segments $M$ (keep in mind that $U_j^{-1} R_f T R_f\, U_j = R_f T R_f$ is always valid for any unitary operator $U_j$, since the orthogonalization action is basis invariant). Therefore, at any point along the backward pass through the fiber, the polarization state is always orthogonal to the state in the forward pass. This time-reversal effect of the Faraday mirror on the polarization state is similar to that of the phase conjugate mirror on the optical wavefront [32]. The scheme does compensate for all polarization perturbations of either mechanical or thermal origins, provided that they are slower than the time of a round trip. It also cancels out all polarization-dependent effects such as polarization-dependent loss and polarization-mode dispersion.

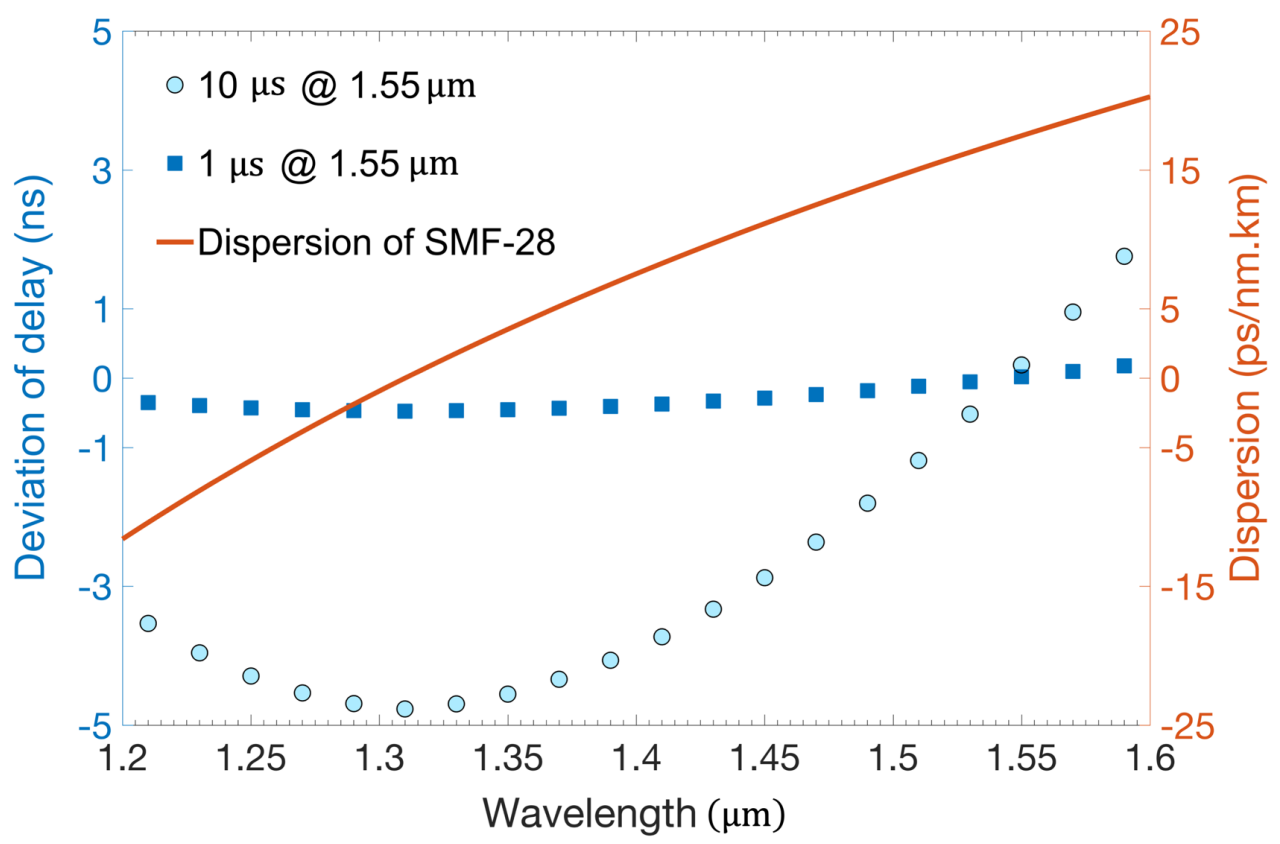

Fig. 3. Wavelength dependence of the delay deviations in cases of 1-μs delay (squares, left axis) and 10-μs delay (circles, left axis), and the dispersion of SMF-28 fiber (solid line, right axis). The deviation of delay is calculated in reference to a delay of 1 μs and 10 μs, precisely realized at 1.55 μm. The spectrally flat performance of the 1-μs delay line is due to the short traverse distance (~200 m) of the light pulse in the optical fiber.

## IV. Broadband 8-Stage 1-$\mu$s Delay Line

### A. Design parameters

Consider a delay line using the commercial SSM fiber "Corning SMF-28", which features effective group indices 1.4673 and 1.4680 at the wavelengths 1310 nm and 1550 nm, respectively [39]. From (3), the maximum delay is $T_{2^K} = (2^K - 1) \times$ resolution step (up to a fixed offset $2K\tau_o$). The resolution step for 8-stage 1-μs delay line is therefore, $2\tau =$ 1 μs / $(2^8 - 1)$ = 3.92 ns.

Based on (1), the system requires a differential length in the first delay stage: $L = \mathrm{v}_g\tau \approx 0.4$ m. The length of the short arm can be negligibly short (e.g., $L_o \approx 0.1$ m). From (3), the maximum propagation distance in the SSM fiber in a round trip is 205.8 m. This relatively short propagation distance helps limit distortion effects due to chromatic dispersion to negligibly small values, as will be seen next.

### B. Propagation in dispersive optical fiber

The dispersion of SMF-28 is determined from the empirical curve-fitting equation

$$D(\lambda) = \frac{S_0}{4}\left(\lambda - \frac{\lambda_0^4}{\lambda^3}\right), \tag{9}$$

where $D(\lambda)$ is the dispersion coefficient at wavelength $\lambda$, measured in ps/nm-km, $S_0$ is the zero-dispersion slope, and $\lambda_0$ is the zero-dispersion wavelength. For SMF-28, $S_0 \leq 0.92$ ps/nm$^2$.km and $\lambda_0 \approx 1310$ nm [39].

As depicted in Fig. 3, the 1-μs delay line exhibits an almost flat spectral performance. The line realizes the 1-μs delay at 1550 nm with a maximum delay deviation < 0.59 ns over the range 1.2–1.6 μm. For a 10-μs delay line, the delay deviation is greater by one order of magnitude, which may require terminal passive or active compensation.

### C. Dispersive Faraday rotation

The wavelength dependence of the Faraday rotation angle affects the polarization-preserving feature of the delay line. The polarization of light traversing the magneto-optical material of the Faraday rotator is rotated by an angle

$$\theta_f(\lambda) = V(\lambda) \int_{-L/2}^{L/2} B(l)\, dl, \tag{10}$$

where $B(l)$ is the effective magnetic field applied along a length $l$, and $V(\lambda)$ is the Verdet constant. The wavelength dependence of the Verdet constant is described by the simplified single-transition model [40, 41]

$$V(\lambda) = \frac{A}{\lambda^2 - \lambda_0^2}, \tag{11}$$

where $\lambda_0$ is a transition wavelength and $A$ is a proportionality factor (both values are determined from experimental measurements).

Considering the wavelength dependence of the rotation angle, the Faraday rotation operator in (5) can be rewritten as

$$R_f(\lambda) = \begin{bmatrix} \exp\left(-i\theta_f(\lambda)\right) & 0 \\ 0 & \exp\left(i\theta_f(\lambda)\right) \end{bmatrix}, \tag{12}$$

which reduces to (5) only at the central wavelength of the Faraday rotator. Consequently, as the wavelength of the optical signal deviates further from central wavelength of the Faraday rotator, the polarization state after the FM will deviate from the condition of orthogonality.

To demonstrate this effect, we consider a Faraday rotator using the CeF3 magneto-optical crystal, which features a high Verdet constant and high transmissivity in the near-infrared (NIR) range [42]. The rotator is designed to apply a 45° rotation at a central wavelength 1.55 μm (corresponding to 90° rotation on the Poincaré sphere). For the CeF3 rotator, the value of the transition wavelength $\lambda_0 = 0.3751$ μm was determined in [42] as the best fit to experimental measurements above 1.1 μm. This assigns a value for the product: $A\int_{-L/2}^{L/2} B(l)\, dl = 1.7764$ rad.μm$^2$.

Figure 4(a) depicts the wavelength dependence of the Faraday rotation angle $\theta_f(\lambda)$ as given by (10) and (11). The operational range for on-shelf NIR Faraday rotators is typically 100 nm. Near the 45° rotation designed at 1.55 $\mu$m, the rotation angle of CeF3 crystal deviates by an angle less than $\Delta\theta_f \approx$ 6.19° in the 1.5 μm – 1.6 μm range. This value accumulates after the FM and is doubled, as shown in Fig. 4(b). Based on (8) and (12), the polarization-preserving property will not be precisely satisfied at wavelengths further from the central wavelength.

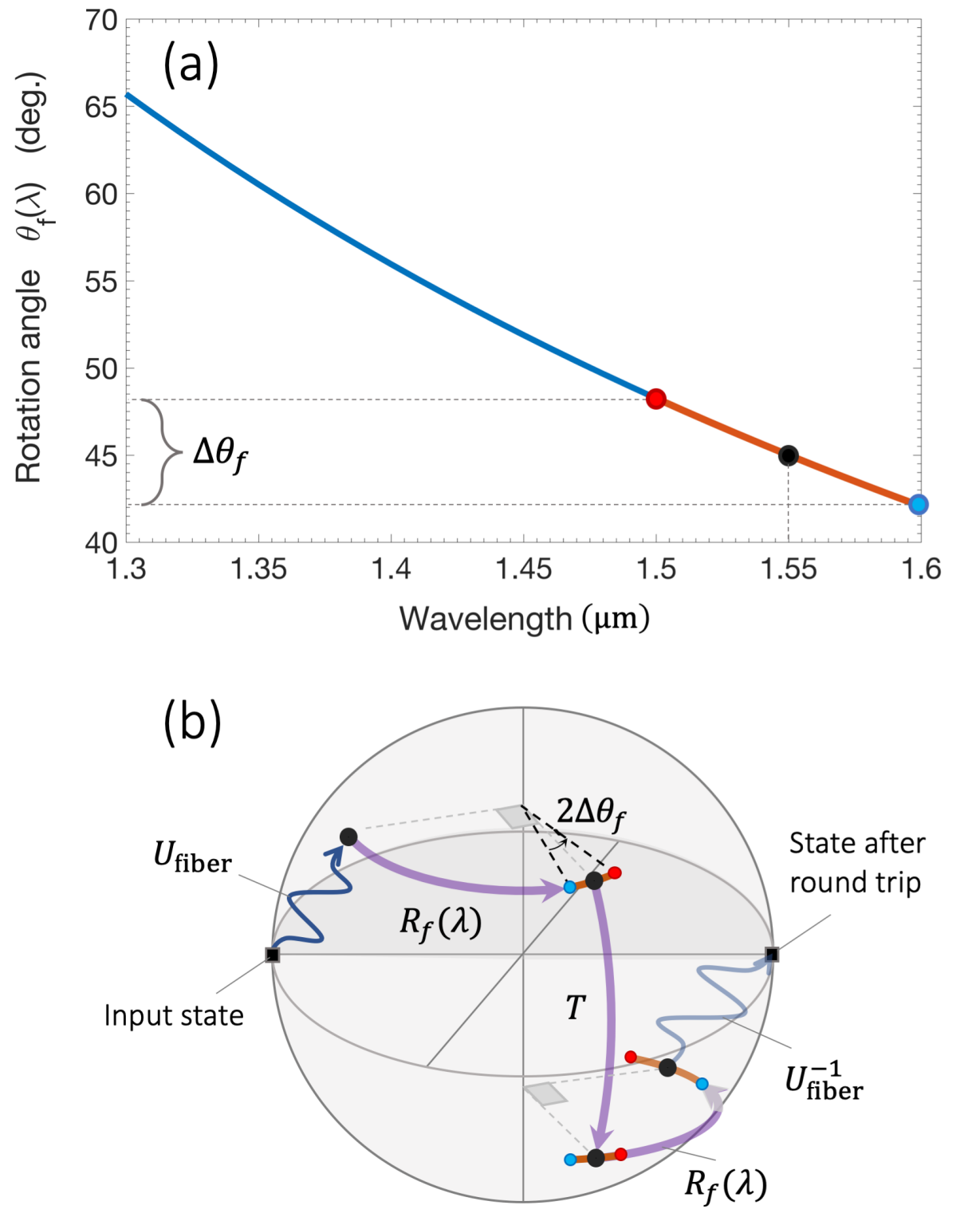

Fig. 4. (a) Wavelength dependence of the rotation angle $\theta_f(\lambda)$ of a $CeF_3$-Faraday rotator designed such that $\theta_f(1.55\ \mu\text{m}) = 45°$. (b) Evolution of the polarization-state on the Poincaré sphere following dispersive Faraday rotation [operator $R_f(\lambda)$], mirror reflection ($T$), and first pass ($U_{\text{fiber}}$) and second pass ($U_{\text{fiber}}^{-1}$) through the optical fiber. The evolved states at 1.5 μm, 1.55 μm, and 1.6 μm are denoted by red, black, and blue circles, respectively. At 1.55 μm, the state after a round trip is always orthogonal to the input state.

### *D. Wavelength-dependent polarization distortion*

We have carried out several numerical experiments to determine the effect of the dispersive Faraday rotation on the polarization-preserving property, and we found that it depends on the polar angle of the input polarization state, but not on its azimuthal angle. We show here results of four numerical experiments. In each experiment, the input polarization state $|\psi_{in}\rangle$ was one of the states:

i. $|H\rangle \equiv (|R\rangle + |L\rangle)/\sqrt{2}$,

ii. $\sqrt{2/3}|R\rangle + \sqrt{1/3}|L\rangle$,

iii. $\sqrt{9/10}|R\rangle + \sqrt{1/10}|L\rangle$,

iv. $|R\rangle$. (13)

All numerical experiments involved a forward pass through the optical fiber followed by reflection from the first FM, backward fiber pass, and finally, reflection from the second identical FM. Every experiment includes 400 runs executed at each of the wavelengths: 1.5 μm, 1.55 μm, and 1.6 μm. For every run, the state after the roundtrip pass through the fiber stages

$$|\psi_{\perp}\rangle = U_{fiber}^{-1}\, R_f(\lambda)\, T\, R_f(\lambda)\, U_{fiber}\, |\psi_{in}\rangle, \quad (14)$$

and the output state of the delay line

$$|\psi_{out}\rangle = R_f(\lambda)\, T\, R_f(\lambda)\ |\psi_{\perp}\rangle, \quad (15)$$

were calculated, where $U_{\text{fiber}}$ is a random unitary matrix representing evolution of the polarization state in the optical fiber,

$$U_r = \begin{bmatrix} a & \exp(i2\pi b)\sqrt{1-a^2} \\ -\exp[i2\pi(c-b)]\sqrt{1-a^2} & a\ \exp(i2\pi c) \end{bmatrix}, \quad (16)$$

and the coefficients $a, b, c$ are statistically independent real random numbers in the [0,1] range.

The resultant state $|\psi_{\perp}\rangle$ in the four experiments is depicted in Fig. 5. As expected, the polarization states for signals at the central wavelength 1.55 μm (denoted by black circles) are always orthogonal to the input polarization states. To the contrary, the output states at 1.5 μm and 1.6 μm (denoted by red and blue dots, respectively) are scattered near the orthogonal state. When $U_{\text{fiber}} = I$, i. e., the optical fiber is eliminated, the output states are denoted by red and blue circles for 1.5 μm and 1.6 $\mu$m, respectively.

Figure 6 shows the results of the four experiments for the output state $|\psi_{\text{out}}\rangle$. At the central wavelength 1.55 μm, the output state always coincides with the input polarization states (denoted by black circles). The output states at 1.5 μm and 1.6 μm (denoted by red and blue dots, respectively) are scattered and dispersed near the input polarization state.

One can notice that over 400 random and independent settings of $U_{\text{fiber}}$, the spread of the scattered states in Figs. 5 and 6 is not centered about the no-fiber case: $U_{\text{fiber}} = I$. Actually, in most of the shown cases in Fig. 6, it is evident that the randomness of $U_{\text{fiber}}$ leads to output states that are often closer to the input state, compared with $U_{\text{fiber}} = I$. Surprisingly, the optical-fiber operator seems to mitigate the effect of the dispersive Faraday rotation on the output state.

This mitigation behavior is quantitatively represented in Fig. 7(a) and 7(b) for the wavelengths 1.5 μm and 1.6 μm, respectively. The fidelity of the output states is numerically calculated for every random setting, $U_{\text{fiber}} = U_r$. For the case without the optical fiber ($U_{\text{fiber}} = I$), the fidelity of the output state is obtained based on the Poincaré-sphere geometry as

$$F(|\psi_{\text{in}}\rangle; |\psi_{\text{out}}\rangle) = \langle\psi_{\text{in}}|\psi_{\text{out}}\rangle$$

$$= \cos\{\sin^{-1}[\sin(4\theta_f(\lambda) - \pi)\sin\theta]\}, \quad (17)$$

where $\theta$ is the polar angle of the input state $|\psi_{\text{in}}\rangle$. It is noted that the mean fidelity for a random settings of the fiber, $U_{\text{fiber}} = U_r$, is higher compared to the no-fiber case $U_{\text{fiber}} = I$ over a wide range of polar angles.

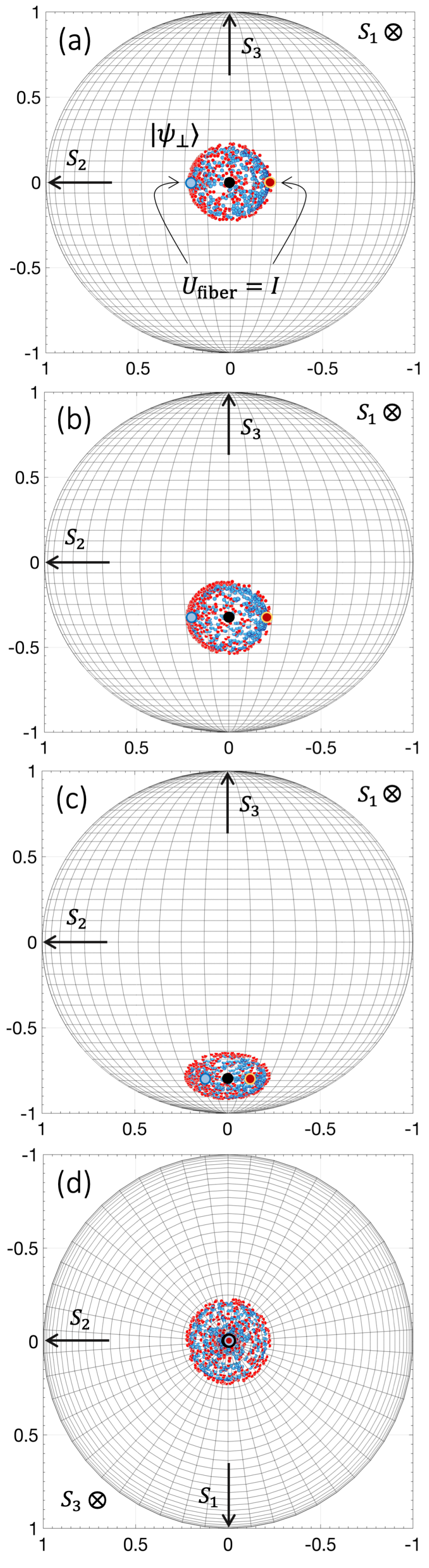

Fig. 5. Results of four numerical experiments showing the polarization state $|\psi_\perp\rangle$ after the bidirectional pass in the fiber stages. The scattered red and blue dots denote polarization states of 400 runs per experiment at 1.5 µm and 1.6 µm, respectively. Every run considers an optical fiber modeled by a random and independent operator as in (16). The red and blue circles denote the output states when $U_{\text{fiber}} = I$. The black circle denotes the state $|\psi_\perp\rangle$ at 1.55 µm, coinciding with the non-distorted state, orthogonal to $|\psi_{\text{in}}\rangle$. Results are shown for (a) $|\psi_{\text{in}}\rangle = |H\rangle \equiv (|R\rangle + |L\rangle)/\sqrt{2}$, (b) $|\psi_{\text{in}}\rangle = \sqrt{2/3}|R\rangle + \sqrt{1/3}|L\rangle$, (c) $|\psi_{\text{in}}\rangle = \sqrt{9/10}|R\rangle + \sqrt{1/10}|L\rangle$, (d) $|\psi_{\text{in}}\rangle = |R\rangle$.

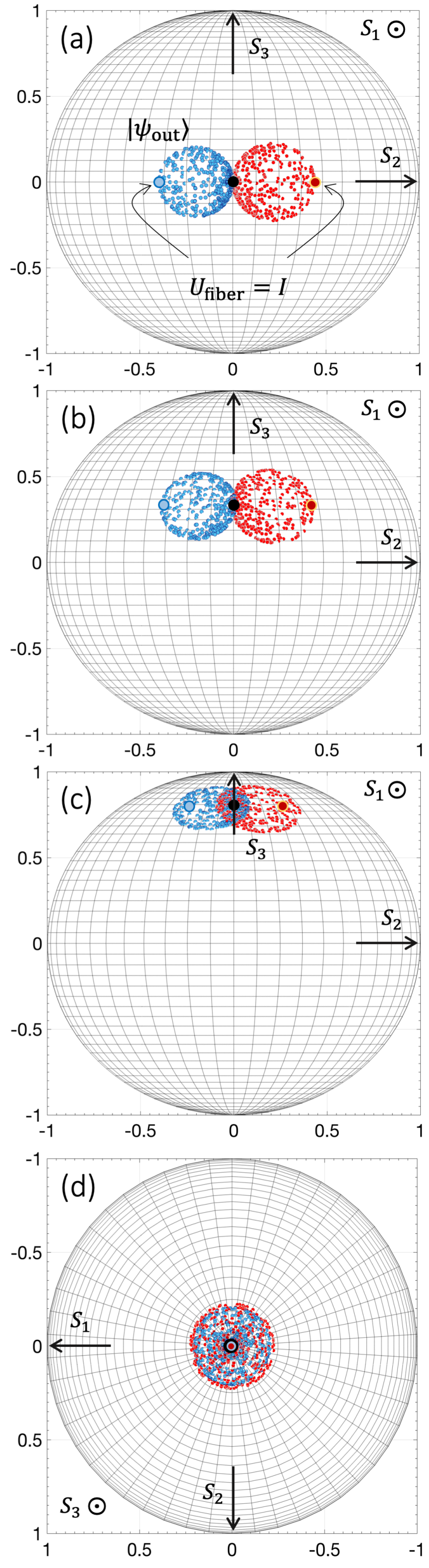

Fig. 6 Distortion of the output polarization state $|\psi_{\text{out}}\rangle$ after the optical delay line. This distortion is due to the interplay between the dispersive Faraday rotation in the $CeF_3$ crystal and the random operator of the optical fiber. The scattered red and blue dots denote output polarization states of 400 runs per experiment at 1.5 µm and 1.6 µm, respectively. The red and blue circles denote the output states when $U_{\text{fiber}} = I$. The black circle denotes the input state, coinciding with the non-distorted output state at 1.55 µm. Results are shown for (a) $|\psi_{\text{in}}\rangle = |H\rangle \equiv (|R\rangle + |L\rangle)/\sqrt{2}$, (b) $|\psi_{\text{in}}\rangle = \sqrt{2/3}|R\rangle + \sqrt{1/3}|L\rangle$, (c) $|\psi_{\text{in}}\rangle = \sqrt{9/10}|R\rangle + \sqrt{1/10}|L\rangle$, (d) $|\psi_{\text{in}}\rangle = |R\rangle$.

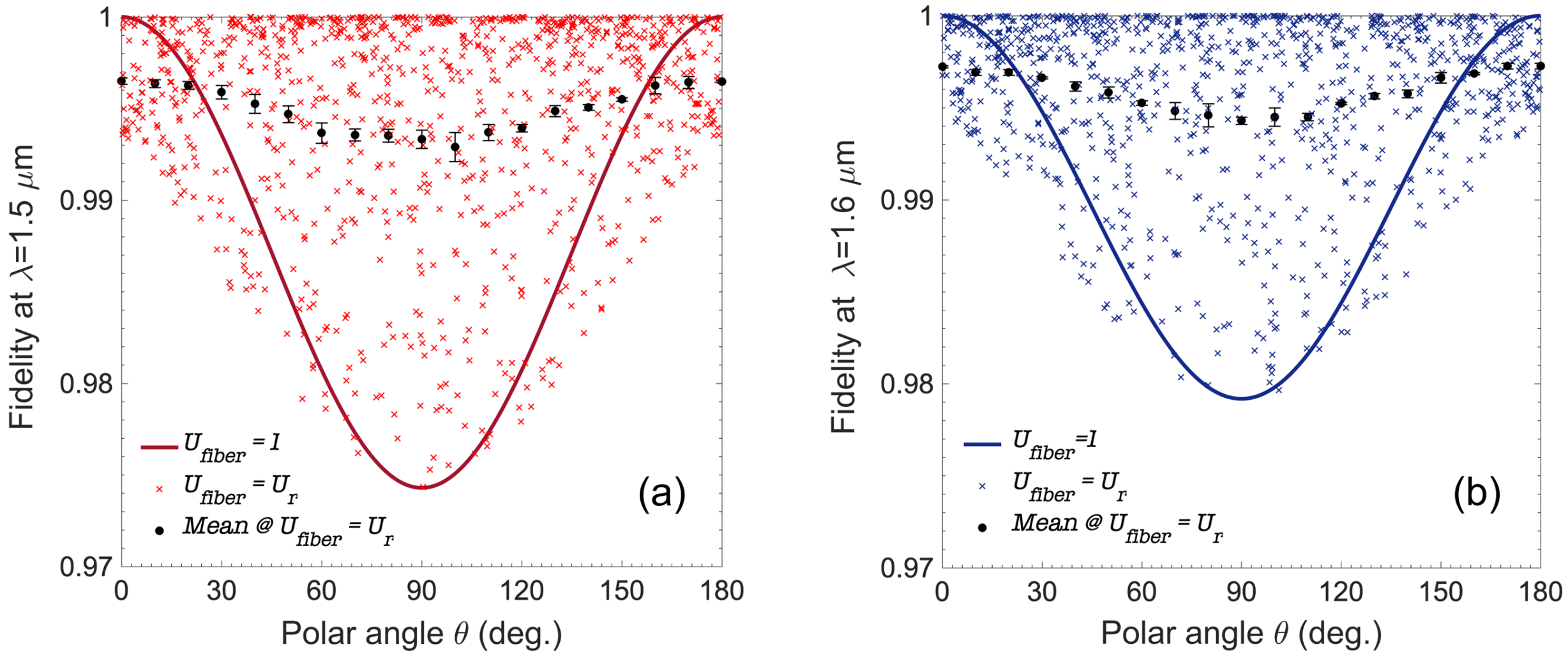

Fig. 7. Fidelity of the output polarization state $|\psi_{\text{out}}\rangle$ to the input state $|\psi_{\text{in}}\rangle$, plotted versus the polar angle of $|\psi_{\text{in}}\rangle$ on the Poincaré sphere, where: (a) $\lambda =$ 1.5 µm, (b) $\lambda =$ 1.6 µm. The optical fiber operator is either substituted by $U_{\text{fiber}} = I$ (no fiber), or randomly set, i.e., $U_{\text{fiber}} = U_r$. The fidelity of the former is obtained from (17); for the latter, it was determined numerically for every random setting of $U_r$ (1000 runs). The mean and standard deviation of the fidelity are calculated for random settings of $U_{\text{fiber}}$ at 10-degree steps of the polar angle $\theta$. Over a wide range of polar angles, the mean fidelity surpasses the no-fiber fidelity ($U_{\text{fiber}} = I$). This shows that the random optical-fiber operator mitigates the effect of the dispersive Faraday rotator on the output state.

The explanation of how the random setting of the optical fiber mitigates the dispersive Faraday rotation is quite simple. In (14) and (15), the input state $|\psi_{\text{in}}\rangle$ is subject to two cascaded dispersive transformations, $U_{\text{fiber}}^{-1}\, R_f(\lambda)\, T\, R_f(\lambda)\, U_{\text{fiber}}$ and $R_f(\lambda)\, T\, R_f(\lambda)$. The unitary fiber transformation can be expressed as

$$U_{\text{fiber}} = \exp(-i\kappa\, \hat{n}.\vec{\sigma}/2), \tag{18}$$

which denotes rotation by angle κ about a unit vector $\hat{n} = (n_x, n_y, n_z)$, where $\vec{\sigma} = (\sigma_x, \sigma_y, \sigma_z)$ is the vector of Pauli matrices. In the transformation $U_{\text{fiber}}^{-1}\, R_f(\lambda)\, T\, R_f(\lambda)\, U_{\text{fiber}}$, the operator $U_{\text{fiber}}$ replaces the z axis of the Poincaré sphere (about which the rotation of $R_f(\lambda)\, T\, R_f(\lambda)$ applies) by another axis. Therefore, while the dispersion of $R_f(\lambda)\, T\, R_f(\lambda)$ introduces a shift in the rotation angle about the $z$ axis, the dispersion of $U_{\text{fiber}}^{-1}\, R_f(\lambda)\, T\, R_f(\lambda)\, U_{\text{fiber}}$ introduces a spread in the rotation angle about that new axis. Because $\hat{n}$ can be any unit vector, the dispersion of $U_{\text{fiber}}^{-1}\, R_f(\lambda)\, T\, R_f(\lambda)\, U_{\text{fiber}}$ will lead to a spread within a circular region, symmetrical around a state orthogonal to the input state (as shown in Fig. 5). The symmetry is due to the random walk by the operator $U_{\text{fiber}}$ in the numerical experiments. Since rotations conserve the Euclidean distance on the Poincaré sphere, the spread is included within a half opening angle $|4\theta_f(\lambda) - \pi|$. On the other hand, $R_f(\lambda)\, T\, R_f(\lambda)$ introduces a difference in the azimuthal angle with the value $[4\theta_f(\lambda) - \pi]$, which can be positive or negative based on the value of λ. Therefore, we observe in Fig. 6 a circular spread of scattering that is centered apart from the initial state by an azimuthal shift $[4\theta_f(\lambda) - \pi]$.

We now consider a case where the optical fiber operator can entirely compensate for the dispersive Faraday rotation. If $\hat{n} = (n_x, n_y, 0)$ and $\theta = \pi$ ($U_{\text{fiber}} = \sigma_x, \sigma_y$ are elements of this set), then $U_{\text{fiber}}$ reverses the $z$ axis of the Poincaré sphere. In this case, the transformation $R_f(\lambda)\, T\, R_f(\lambda)$ will precisely do the opposite action of $U_{\text{fiber}}^{-1}\, R_f(\lambda)\, T\, R_f(\lambda)\, U_{\text{fiber}}$. The two transformations involve right-handed and left-handed rotations by the same dispersive angle about the z axis. This cancels out the dispersive effect, and the initial state is restored. The output state in this case coincides with the initial state (denoted by black circles in Fig. 6).

## V. Discussion and Conclusion

We have introduced a broadband programmable binary optical delay line that preserves the polarization state of the delayed signal. The delay line uses a double-pass single-mode fiber with two terminal Faraday mirrors that help eliminate random polarization fluctuations introduced by the optical fiber, thereby restoring the initial polarization state. We assessed the effect of the dispersive character of practical Faraday rotators on the bandwidth of the delay line. The results reveal a surprising behavior when the optical fiber is sandwiched between two terminal Faraday mirrors— random polarization distortion due to fiber propagation remedies FM dispersion and sustains the polarization-preserving property over a wide bandwidth. While the experimental results are demonstrated for the $CeF_3$ FM, the broadband polarization preservation realized by our system is applicable to other magneto-optic crystals.

The configuration of the delay line presented in this paper relies on optical switches that choose between two-path modes of different lengths. Another implementation may consider the differential group delay (DGD) between two polarization modes of the polarization maintaining (PM) fiber [5,7]. In this implementation, the optical switch and two SSM fiber segments in each delay stage are replaced by a polarization switch inserted into a PM fiber. However, it should be noted that the PM fiber cannot maintain an arbitrary polarization state. It only maintains the polarization of its two principal states, and other states will undergo fast evolution due to the large birefringence of the principal polarization modes. Also, the DGD has a remarkable spectral behavior that renders the delay line highly dispersive. In addition, because of the limited group-index birefringence in practice, it requires a relatively long PM fiber (much longer than the SSM fiber) in order to achieve a 1-μs delay. Taking into account that the PM fiber has significantly higher absorption coefficient than that of the SSM fiber, this implementation renders the optical signal highly subject to loss, and is therefore more appropriate for use as a few-ns delay line.

## REFERENCES

[1] G. Daniel, "Slow light brings faster communications," *Phys. World* 18, pp. 30–32, 2005.
[2] J. T. Mok and B. J. Eggleton, "Photonics: Expect more delays," *Nature*, vol. 433, pp. 811–812, 2005.
[3] L. Zhuang *et al.*, "Novel low-loss waveguide delay lines using vernier ring resonators for on-chip multi-λ microwave photonic signal processors," *Laser Photon. Rev.*, vol. 7, pp. 994–1002, 2013.
[4] D. R. Gervais, J. R. Minch, and D. J. Townsend. "Rate adjustable differential phase shift key (DPSK) modulation." U.S. Patent No. 7,817,924, 2010.
[5] A. E., Morra *et al.*, "Hybrid direct-detection differential phase shift keying-multipulse pulse position modulation techniques for optical communication systems," *Opt. Commun.*, vol. 357, pp. 86-94, 2015.
[6] A. E. Al-Fiqi *et al.*, "Performance Evaluation of Hybrid DPSK-MPPM Techniques in Long-Haul Optical Transmission," *Appl. Opt.*, vol. 55, no. 21, pp. 5614-5622, 2016.
[7] A. E. Morra *et al.*, "Hybrid Direct-Detection Differential Phase Shift Keying-Multipulse Pulse Position Modulation Techniques for Optical Communication Systems," US Patent 20180069636A1, 2018.
[8] A. E. Al-Fiqi *et al.*, "Chaotic Polarization-Assisted LDPSK-MPPM Modulation for Free-Space Optical Communications," *IEEE Transactions on Wireless Communications*, vol. 18, no. 9, pp. 4225 – 4237, 2019.
[9] A. Elsonbaty, S. F. Hegazy, and S. S. A. Obayya "Simultaneous Suppression of Time-Delay Signature in Intensity and Phase of Dual-Channel Chaos Communication," *IEEE Journal of Quantum Electronics*, vol. 51, no. 9, pp. 1-9, 2015; A. Elsonbaty, S. F. Hegazy, and S. S. A. Obayya, "Simultaneous concealment of time delay signature in chaotic nanolaser with hybrid feedback," *J. Opt. Lasers Eng.*, vol. 107, pp. 342–351, 2018.
[10] A. Elsonbaty, S. F. Hegazy, and S. S. A. Obayya "Numerical analysis of ultrafast physical random number generator using dual-channel optical chaos," *Opt. Engin.*, vol. 55, no. 9, pp. 094105-094105, 2016.
[11] F. Kaneda *et al.*, "Quantum-memory-assisted multi-photon generation for efficient quantum information processing." *Optica*, vol. 4, no. 9, pp. 1034-1037, 2017.
[12] S. F. Hegazy and S. S. A. Obayya, "Tunable spatial–spectral phase compensation of type-I (ooe) hyperentangled photons," *J. Opt. Soc. Am. B,* vol. 32, no. 3, pp. 445-450, 2015; S. F. Hegazy, Y. A. Badr, and S. S. A. Obayya, "Relative-phase and time-delay maps all over the emission cone of hyperentangled photon source," *Opt. Eng.*, vol. 56, no. 2, pp. 026114-026114, 2017.
[13] S. F. Hegazy, S. S. A. Obayya, and B. E. A. Saleh, "Orthogonal quasi-phase-matched superlattice for generation of hyperentangled photons," *Sci. Rep.*, vol. 7, 4169, 2017; S. F. Hegazy and S. S. A. Obayya, "Extended Source of Indistinguishable Polarization-entangled Photons over Wide Angles of Emission," *J. Appl. Lett.*, vol. 117, no. 24, 244003, 2020.
[14] N. Gisin, G. Ribordy, W. Tittel, and H. Zbinden,"Quantum cryptography," *Rev. Mod. Phys.*, vol. 74, pp. 145–195, 2002.
[15] D. Stucki *et al.*, "Quantum key distribution over 67 km with a plug & play system," *New J. Phys.*, vol. 4, pp. 41.1–41.8, 2002.
[16] S. F. Hegazy, S. S. A. Obayya, and B. E. A. Saleh, "Randomized ancillary qubit overcomes detector-control and intercept-resend hacking of quantum key distribution," *IEEE J. Lightwave Technol.*, vol. 40, no. 21, pp. 6995-7005, 2022.
[17] F. Xu, X. Ma, Q. Zhang, H. K. Lo, and J. W. Pan, "Secure quantum key distribution with realistic devices," *Rev. Mod. Phys.*, vol. 92, Art. no. 025002, 2020.
[18] I. Giuntoni *et al.*, "Continuously tunable delay line based on SOI tapered Bragg gratings," *Opt. Express,* vol. 20, pp. 11241–11246, 2012.
[19] Y. A. Vlasov, M. O'Boyle, H. F. Hamann, and S. J. McNab, "Active control of slow light on a chip with photonic crystal waveguides," *Nature*, vol. 438, pp. 65–69, 2005.
[20] D. O'Brien *et al.*, "Tunable optical delay using photonic crystal heterostructure nanocavities," *Phys. Rev. B*, vol. 76, Art. no. 115110, 2007.
[21] J. Sancho *et al.*, "Integrable microwave filter based on a photonic crystal delay line," *Nat. Commun.*, vol. 3, no. 1, Art. no. 1075, 2012.
[22] N. Fontaine *et al.*, "Continuously tunable optical buffering at 40 gb/s for optical packet switching networks," J. Lightwave Technol. vol. 26, pp. 3776–3783, 2008.
[23] J. Yang *et al.*, "Continuously tunable, wavelength-selective buffering in optical packet switching networks," *IEEE Photon. Technol. Lett.*, vol. 20, pp. 1030–1032, 2008.
[24] Y. Okawachi *et al.*, "All-optical slow-light on a photonic chip," *Opt. Express*, vol. 14, pp. 2317–2322, 2006.
[25] F. Morichetti *et al.*, "A reconfigurable architecture for continuously variable optical slow-wave delay lines," *Opt. Express*, vol. 15, pp. 17273–17282, 2007.
[26] J. B. Khurgin, "Dispersion and loss limitations on the performance of optical delay lines based on coupled resonant structures," *Opt. Lett.*, Vol 32, pp. 133–135, 2007.
[27] D. Dolfi *et al.*, "Experimental demonstration of a phased-array antenna optically controlled with phase and time delays," *Appl. Opt.*, vol. 35, pp. 5293–5300, 1996.
[28] K. Jinguji, N. Takato, Y. Hida, T. Kitoh, and M. Kawachi,"Two-port optical wavelength circuits composed of cascaded Mach-Zehnder interferometers with point-symmetrical configurations," *J. Lightwave Technol.,* vol.14, pp. 2301–2310, 1996.
[29] S. Li, X. Li, W. Zou, and J. Chen, "Rangeability extension of fiber-optic low-coherence measurement based on cascaded multistage fiber delay line," *Appl. Opt.*, vol. 51, pp. 771–775, 2012.
[30] X. Wang *et al.*, "Continuously tunable ultra-thin silicon waveguide optical delay line." *Optica*, vol. 4, no. 5, pp. 507-515, 2017.
[31] M. Martinelli, ''A universal compensator for polarization changes induced by birefringence on a retracing beam,'' *Opt. Commun.*, vol. 72, pp. 341–344, 1989.
[32] M. Martinelli, ''Time reversal for the polarization state in optical systems,'' *J. Mod. Opt.*, vol. 39, pp. 451–455, 1992.
[33] M. Martinelli and P. Martelli, "Polarization, mirrors, and reciprocity: birefringence and its compensation in optical retracing circuits," *Adv. Opt. Photon.*, vol. 9, no. 1, pp. 129-168, 2017.
[34] M. Martinelli, L. Marazzi, P. Parolari, M. Brunero, and G. Gavioli, "Polarization in retracing circuits for WDM-PON," *IEEE Photon. Technol. Lett.*, vol. 24, pp. 1191–1193, 2012.
[35] P. Parolari *et al.*, "10 Gb/s operation of a color- less self-seeded transmitter over more than 70 km of SSMF," *IEEE Photon. Technol. Lett.*, vol. 26, pp. 599–602, 2014.
[36] A. D. Kersey, M. J. Marrone, and M. A. Davis, "Polarization-insensitive fibre optic Michelson interferometer," *Electron. Lett.*, vol. 27, pp. 518–520, 1991.
[37] N. A. Riza and N. Madamopoulos, "High signal-to-noise ratio birefringence-compensated optical delay line based on a noise-reduction scheme," *Optics Letters*, vol. 20, no. 22, p. 2351,1995.
[38] N. A. Riza and N. Madamopoulos, "Microwave band demonstration of a reflective geometry fiber and free-space binary photonic delay line," *IEEE Microwave and Guided Wave Letters*, vol. 7, no. 4, pp. 103–105, 1997.
[39] https://www.corning.com/content/dam/corning/media/worldwide/coc/documents/Fiber/PI-1500.pdf
[40] J. C. Suits, B. E. Argyle, and M. J. Freiser, "Magneto-Optical properties of materials containing divalent europium," *J. Appl. Phys.*, vol. 37, no. 3, pp. 1391–1397, 1966.
[41] V. Vasyliev, E. G. Víllora, M. Nakamura, Y. Sugahara, and K. Shimamura, "UV-visible Faraday rotators based on rare-earth fluoride single crystals: LiREF_4 (RE = Tb, Dy, Ho, Er and Yb), PrF_3 and CeF_3," *Opt. Express*, vol. 20, no. 13, p. 14460, 2012.

[42] D. Vojna *et al.*, "Verdet constant dispersion of $CeF_3$ in the visible and near-infrared spectral range," *Opt. Eng.*, vol. 56, no. 6, p. 067105, 2017.

**Salem F. Hegazy** (Member, IEEE) was born in Hulwan, Egypt. He received the B.Sc. (Hons.), Diploma, and M.Sc. degrees in electronics and communications engineering from Cairo University, Giza, Egypt, in 2002, 2006, and 2011, respectively, and the Ph.D. degree in quantum optics and nonlinear optics under a joint-supervision scheme between CREOL, The College of Optics and Photonics, University of Central Florida, Orlando, FL, USA, and the National Institute of Laser Enhanced Sciences (NILES), Cairo University, in 2017. He has been an Associate Professor with the National Institute of Laser Enhanced Sciences, Cairo University, and also a Senior Research Fellow with the Centre for Photonics and Smart Materials, Zewail City of Science and Technology, Giza, Egypt. His research interests include optical quantum-information and quantum- computation processing, practical security of quantum cryptography, photonic entanglement, optical communications, and optical chaos-based systems. He was the recipient of the Egyptian State Encouragement Award in Engineering Sciences in 2019 presented to distinguished scholars under 40 years.

**Salah S. A. Obayya** (Fellow, IEEE) received the B.Sc. (Hons.) and M.Sc. degrees in electronics and communications engineering and the Ph.D. degree from Mansoura University, Mansoura, Egypt, in 1991, 1994, and 1999, respectively. From 1997 to 1999, he was with the Department of Electrical, Electronic, and Information Engineering, City University London, London, U.K., to carry out the research part of his Ph.D. under the joint supervision scheme between the City University London and Mansoura University. From 2000 to 2003, he was a Senior Research Fellow with the School of Engineering, City University London. From 2003 to 2006, he was with the School of Engineering and Design, Brunel University, London, U.K., as a Senior Lecturer, and a Reader with the School of Electronic and Electrical Engineering, University of Leeds, Leeds, U.K. Since 2008, he has been a Full Professorial Chair of photonics and led the establish- ment of the Nano-Photonics Research Centre, The University of Glamorgan, Pontypridd, U.K., while he was currently the Vice Chair for Academic Affairs, Zewail City of Science and Technology, Giza, Egypt. He has built an outstanding international reputation in the area of green nanophotonics with a focus on the intelligent computational modeling of modern nanophotonic devices enabling technologies for efficient generation, distribution, and utilization of sustainable energy toward low-carbon green society. He has a track record of 232 journal publications mostly in the IEEE, IET/IEE, and OSA, has authored three books, namely *Computational Photonics* (Wiley, 2010), *Computational Liquid Crystal Photonics* (Wiley, 2016), and Computational of Photonic Sensors (2018), which are adopted references for Graduate Programs in Photonics at international Universities, and 252 conference papers in international conference in photonics, many of these are invited talks, and keynote lectures, attracted external funding, from both industry and Research Councils, in the region of $4 M, and supervised to successful completion 35 Ph.D. students, over 70 M.Sc./M.Res./M.Phil. students, and 15 postdoctoral research fellows.

**Bahaa E. A. Saleh** (Life Fellow, IEEE) received the Ph.D. degree from the Johns Hopkins University, Baltimore, MD, USA, in 1971, He has been Distinguished Professor with CREOL, The College of Optics and Photonics, University of Central Florida, Orlando, FL, USA, since 2009, and he was the Dean in 2009–2019. He held faculty and research positions with the University of Santa Catarina, Florianópolis, Brazil, Max Planck Institute, Germany, University of California-Berkeley, Berkeley, CA, USA, Columbia University, New York, NY, USA, University of Wisconsin-Madison, Madison, WI, USA, and Boston University, Boston, MA, USA, where he was the Chair of the ECE Department (1994–2008). He has made significant contributions in areas including coherence and statistical optics, nonlinear and quantum optics, and image science. His pub- lications include more than 285 journal papers and three books: *Photoelectron Statistics*, (Springer, 1978), *Fundamentals of Photonics* (Wiley, 2007, with M. C. Teich), and *Introduction to Subsurface Imaging* (Cambridge, 2011). He was the Editor-in-Chief of *Journal of the Optical Society of America A* (1991–1997) and founding Editor of *OSA Advances in Optics and Photonics* (2008–2013). He is Fellow of OSA, APS, SPIE, and Guggenheim Foundation. He was the recipient of the OSA Beller Award, SPIE BACUS Award, Kuwait Prize, OSA Distinguished Service Award, and OSA Mees Medal.